\documentclass{JHEP3}

\usepackage{epsfig}

\makeatletter\@addtoreset{equation}{section}\makeatother

\def\be{\begin{equation}}
\def\ee{\end{equation}}
\def\bear{\begin{eqnarray}}
\def\eear{\end{eqnarray}}
\def\nn{\nonumber}

\def\half{{{1\over 2}}}

\def\Im{{\rm Im\hskip0.1em}}

\newcommand\px[1]{{\partial_{#1}}}

\newcommand\Tr[1]{{\mbox{Tr}\{{#1}\}}}          

\def\a{{\alpha}}
\def\b{{\beta}}

\def\lam{{\lambda}}

\preprint{\tt {hep-th/0306132} \\ RUNHETC-2003-18}
\author{Joanna L. Karczmarek$^1$,  Hong Liu$^2$, Juan Maldacena$^3$,
and Andrew Strominger$^1$\\
{\it
$^1$ Jefferson Physical Laboratory,
Harvard University\\
Cambridge, MA, 02138
\vskip 0.15in
$^2$ Department of Physics, Rutgers University \\
Piscataway, NJ, 08855-0849
\vskip 0.15in
$^3$ Institute for Advanced Study\\
Princeton, NJ, 08540
}}
\title{UV Finite Brane Decay }
\abstract{
The decay of an unstable D-brane via  closed string emission and open
string pair production  is considered in subcritical string theory with a
spacelike linear dilaton.  The decay rate is given by
the imaginary part of the annulus, which has  ambiguities corresponding to
the choices of incoming closed and open string vacua.
An exact expression for the full annulus diagram
is computed with a natural choice of incoming vacua. It
is found that the ultraviolet divergences present in critical
string theory in both of these processes are absent for any nonzero
spacelike dilaton. Implications for the vexing issue of
the tachyon dust are
discussed.
}
\keywords{Bosonic Strings, Tachyon Condensation, Conformal Field Models in String Theory}

\begin{document}

\section{Introduction}
\label{sec:intro}
\noindent\indent
The study of s-branes, or time-dependent open string tachyons,
has led to several interesting puzzles. These puzzles are technically
challenging and their resolutions may lead to new conceptual
insights. Recent work includes
\nocite{Gutperle:2002ai, Sen:2002nu,Sen:2002qa,
Strominger:2002pc,Okuda:2002yd, Chen:2002fp,
Sen:2002an, Sen:2002in,  Constable:2003rc, Larsen:2002wc,
Mukhopadhyay:2002en, Maloney:2003ck, Lambert:2003zr, Gaiotto:2003rm}
\cite{Gutperle:2002ai}--\cite{Gaiotto:2003rm}.

The first puzzle is the appearance of the so-called tachyon dust
\cite{Sen:2002nu}.  Open string tachyon condensation, or equivalently brane
decay, can be described by an exact worldsheet CFT. Using this description
Sen \cite{Sen:2002nu} argued that, for $g_s=0$, at the endpoint
of tachyon condensation the unstable brane is replaced by a pressureless
gas of ``tachyon dust''.  This seems peculiar because, once the tachyon fully
condenses, the brane should disappear and one expects no open string
excitations. Rather, all excitations should be described in terms of closed
strings.
This raises the question: Is the tachyon dust closed strings in disguise,
or a new form of matter in string theory?

It was subsequently suggested \cite{Strominger:2002pc} that the
appearance of the tachyon dust is an unphysical artifact of
perturbation theory which does not survive to any finite, nonzero
$g_s$. The time-dependent tachyon background leads to open string
pair production with Hagedorn-like divergences. Hence at any
finite value of $g_s$ perturbation theory will break down in a
time of order one in string units, long before the tachyon dust is
encountered, and the brane energy is presumably converted into
outgoing closed strings. In \cite{Lambert:2003zr} a similar
Hagedorn-like divergence was found in closed string emission, and
was also argued to lead to the breakdown of perturbation theory
and the conversion of all brane energy into outgoing closed
strings. It was further suggested 
\cite{Lambert:2003zr, Sen:2003bc, Shiraz} that the
tachyon dust might actually be comprised of the
highly massive closed strings emitted during the decay
process.

While it may seem comforting that the breakdown of perturbation
theory saves us from a direct confrontation with the tachyon dust,
in this paper we will find that both the open and closed string
divergences are controlled in a subcritical string theory with
a spacelike linear dilaton. The result is finite\footnote{Up to
certain tachyon divergences which are presumably absent in the
full s-brane of the superstring, as discussed in section 4.2.} (per unit spatial volume)
expressions for both the closed string emission and the open
string pair production. The basic reason for this is that
subcritical string theory has a higher Hagedorn temperature.
In this setting
perturbation theory does not seem to break down, and the energy in
closed string emission is insufficient (by a factor of $g_s$) to
account for the energy in the tachyon dust as proposed in
\cite{Lambert:2003zr, Sen:2003bc, Shiraz}.\footnote{However in section 3
we will find some evidence supporting the idea that the tachyon dust is 
composed of massive closed strings (of unknown origin).} The dust typically
drifts into the weak coupling region where it might be perturbatively
studied. Hence the tachyon dust puzzle has been at least temporarily
resuscitated, although we are confident that this paper will not be the
last word on the subject.

The main result of this paper, on which the preceding statements are
based, is the expression for the worldsheet
partition function for unstable for $p$-brane decay in
$D\le 26$-dimensional string theory \be
 Z\left(\frac{is}{2\pi}\right) =\frac{2V_p}{(4\pi s)^{\frac{p+2}{2}}}
\int_{-\infty}^\infty d\nu e^{-s\nu^2}\eta^{2-D}\left(\frac{is}{2\pi}\right)
\partial_\nu \ln S_\beta\left(
\frac{\beta}{2}+\frac{1}{2\beta}-\nu \right),
\label{bst}
\ee
where the
dilaton gradient has a timelike component
$\partial_0\Phi=\beta-\frac{1}{\beta}$
, $S_\beta$ is the
$q$-gamma function, $s$ is the modular parameter of the annulus and
the $\nu$ integration contour is just
above (below) the positive (negative) real axis.
$Z$ acquires an imaginary part from
integrating past the poles given by 
\be \Im {Z}\left(\frac{is}{2\pi}\right) = V_p
\sum_{m,n=0}^{\infty} \frac{\eta^{2-D} \left(
\frac{is}{2\pi}\right) }{s (4\pi s)^{\frac{p}{2}}} ~ e^{-\left((n+\half)
\beta +(m+\half)/ \beta \right)^2 s}~  ~. \label{eqn:Zintro}
\ee
Interpreting the imaginary
part as due to closed string emission, this agrees with
the result of \cite{Lambert:2003zr} for closed string emission
during brane decay, and generalizes it to the case of a linear
dilaton. When $D\to 26$, $\beta \to 1$ and the emission
is divergent. We show that for $D<26$ this divergence is
eliminated.  Though this divergence is gone in all the different
decay channels which we have considered, it is possible that there 
exist other decay channels where the emission rate diverges again.

This result sharpens the tachyon dust puzzle while underscoring a second
puzzle. We are used to the idea that classical solutions of string theory
correspond to worldsheet conformal field theories, which in turn have
well defined partition functions and annulus diagrams. However it
is clear that life cannot be quite so simple in the time dependent
context. Formally the annulus diagram describes both open string pair
production and closed string emission. However these are $not$ determined
by the classical solution alone: the incoming state of both the
open and closed strings must be specified.  Hence there cannot be
a uniquely defined annulus diagram for every
classical solution of string theory. In
practice what happens is that the CFTs which describe time dependent
string backgrounds are a priori ill-defined because the worldsheet fields
have negative kinetic terms. A prescription must be introduced - often some
kind of analytic continuation or contour - for defining them. Hence in practice there
may be more than one set of correlators/CFT corresponding to a given
classical solution.\footnote{An example of this was
recently discussed in \cite{Schomerus:2003vv}.}  It has been suggested
\cite{Gutperle:2003xf}
that this mathematical ambiguity in the CFT corresponds to the
physical ambiguity in the choice of vacua.

At present we have no detailed understanding of how these ambiguities match
up.  The case at hand provides a concrete example in which
this issue can be addressed. The annulus diagram has an imaginary part
which can be identified, in the closed string channel, as coming from
on-shell
closed string emission.  On the other hand,
the contribution in the open string channel
from on-shell open strings is purely real, and there is no obvious signal
of decay via open string pair production.
One possible explanation is that the
prescription adopted herein
for the annulus corresponds to the usual Minkowski vacuum
state for the incoming closed strings,
and to the $out$ vacuum for the
incoming open strings, for which there are no outgoing open strings
and hence no associated imaginary part.\footnote{The Minkowski
closed string vacuum arises naturally in the contour prescription of
\cite{Lambert:2003zr} for half s-branes, while the $out$
vacuum for open strings arises naturally in the
analytic continuation from the positive-norm Liouville theory
\cite{Gutperle:2003xf}.} A second possibility is that the annulus needs a
second term containing the representations corresponding to on-shell open
strings. Yet a third possibility is that the closed string emission has
a dual channel interpretation as open string pair production.
Our calculations give some insight into this problem. These possibilities
do not all seem equally likely but we will
not arrive at a definitive conclusion.

Some interesting insights into these issues have recently been obtained
in the context of $c=1$ string theory \cite{ McGreevy:2003kb,
Klebanov:2003km, McGreevy:2003ep}.

This paper is organized as follows.  In section \ref{sec:ld}, we
review some facts about the physics in the linear dilaton background.
In section \ref{sec:stress} we compute the stress-energy tensor
for a brane decaying in this background, and derive a modified
conservation law.  In section \ref{sec:closed}, we define and
compute both the real
and imaginary parts of the annulus in this background
and show that it is finite (up to
the usual tachyon divergences). We also show in the open
string channel that (\ref{bst}) is real (even after integrating over
the modular parameter) near the massive
on-shell open states where open string pair
production could show up. In section
\ref{sec:open} we compute the asymptotic form of open string pair
production in the $in$ vacuum, and again find a UV finite result
when there is a spacelike linear dilaton.

Throughout this paper, we take $\a'=1$.


\section{Strings in the linear dilaton background}
\label{sec:ld}

In this section, we discuss the physics of free
strings in the linear dilaton background \cite{Myers:1987fv}.

Consider string theory in flat space
with a linear dilaton $\Phi = V_\mu X^\mu$.
The linear dilaton worldsheet CFT differs from the regular free boson
CFT only in the quantum improvement term for the energy momentum
tensor \cite{Polchinski:1998rq}
\bear
T(z) = - \partial X^\mu \partial X_\mu  +V_\mu
\partial^2 X^\mu~, \\
T(\bar z) = -  \bar \partial X^\mu \bar \partial X_\mu
+V_\mu \bar \partial^2 X^\mu~.
\label{eqn:hagedorn:T}
\eear
The central charge is modified to be $c = \tilde c = D + 6V^2$,
where $D$ is the number of spacetime directions and $V^2 \equiv
V_\mu V^\mu$. For nonzero $V$, $D$ must be adjusted to maintain
$c=26$.\footnote{We assume $6 V^2$ is an integer so that this is
possible, although the more general case is easily treated using more
general CFTs.} The mass-shell conditions for closed strings are
\be
-(p + iV)^2 = 2 ( N + \tilde N - 2) + V^2 ~,
\label{eqn:hagedorn:L0}
\ee
together with the level matching condition $N = \tilde N$.
On-shell momenta thus have an imaginary component, $-iV_\mu$.
This is natural when one examines, for example, the Lagrangian
for a closed string mode of mass $m^2=2 (N + \tilde N - 2 )$ in this background
\be
{\cal{L}}[\varphi] = e^{-2V_\mu X^\mu}
\left(-\eta^{\mu\nu} \px \mu \varphi \px \nu \varphi
- m^2 \varphi^2 \right)~.
\label{eqn:Lstringframe}
\ee
Defining a new field, $\tilde \varphi \equiv e^{-V_\mu X^\mu} \varphi$,
we obtain
\be
{\cal L}[\tilde \phi] =  -\eta^{\mu\nu} \px \mu \tilde \varphi
\px \nu \tilde \varphi - (m^2 + V^2) \tilde \varphi^2 + \mbox{total
derivative}~,
\ee
giving $\tilde \varphi$ a mass squared
\be
\tilde m^2= {2} (N + \tilde N - 2 )+V^2~.
\ee
We can refer to $\varphi$ as the string frame field.  The
string frame stress-energy tensor is
\be
T_{\mu\nu} \equiv \frac {\delta S}{\delta g^{\mu\nu}}
= e^{-2\Phi} \left [
\partial_\mu \varphi\partial_\nu \varphi
-\half \eta_{\mu\nu} ((\partial\varphi)^2 + m^2\varphi^2)
\right ]~,
\ee
and it obeys a modified conservation law
\be
\partial^\mu T_{\mu\nu} = -V_\nu {\cal{L}} =
\half V_\nu \frac {\delta S}{\delta \Phi}~.
\label{modified.conservation.intro}
\ee

We will refer to the vertex operators of the from
$e^{i k_\mu X^\mu + V_\mu X^\mu}$ as the dressed
vertex operators.  As seen above, the dressing
is necessary to make the physical field normalizable
with respect to the norm given by the action.

For the open string spectrum the arguments are similar.
However, since the open string action has only one
power of string coupling in front of it, the open
string dressed operators are of the form
$e^{i k_\mu X^\mu + \half V_\mu X^\mu}$.
The mass-shell condition is
\be
-\left (p+\frac{i}{2} V\right)^2 = (N-1) + \frac{1}{4}V^2~.
\ee

Since the number of string oscillators is proportional to $(D-2)$, the
Hagedorn temperature becomes
\be
T_H^{(D)} = \frac{1} {2 \pi \sqrt{ \frac{D-2} {6}}}=
\frac{1} {2 \pi \sqrt{ (4-V^2)}}~.
\label{thag}
\ee
For subcritical string theory (spacelike $V_\mu$), this
is higher than for $V_\mu=0$.


\section{Stress-energy of subcritical brane decay}
\label{sec:stress}

In this section we begin our analysis of a D-brane decaying
in the linear-dilaton background by computing the stress-energy,
generalizing the computation of \cite{Sen:2002nu}.

Tachyon condensates on an unstable brane can be of the general
form $e^{t_\mu X^\mu}$. However for the most part we shall find it
convenient to boost into a frame which brings this into the form
$\exp(\beta X^0)$, with $\beta > 0 $. We want to study this in an
arbitrary linear dilaton of the form $\Phi = V_\mu X^\mu$.  We
require the dilaton gradient to point along the unstable brane
(in order to avoid a force on the brane), which,
if the dilaton is spacelike, is possible only for $p>0$.
The weight of the worldsheet
boundary interaction operator is $\beta(\beta-V_0)$. Requiring
this to be $1$, so the interaction is marginal, we obtain \be
V_0 = \beta - 1/\beta ~.\ee
The timelike
part of the worldsheet CFT is the TBL (Timelike Boundary Liouville) theory
\be -\frac{1}{2\pi} \int_{\Sigma}
\left ( \partial X^0 \bar
\partial X^0 - \frac{V_0X^0 R}{4}  \right ) +\frac{1}{2\pi}
\int_{\partial \Sigma} \left (
 {\pi \lambda} e^{\beta X^0}+{V_0X^0 K }
\right )~,
\label{vko}
\ee
where $K$ is the extrinsic curvature which
integrates to $2 \pi$ around the boundary of a disc.

(\ref{vko}) is related to the standard boundary Liouville theory
with $Q = b + \frac{1}{b}$:
\be \frac{1}{2\pi} \int_{\Sigma} \left ( \partial \phi
\bar \partial \phi + \frac{Q\phi  R}{4}  \right ) +\frac{1}{2\pi}
\int_{\partial \Sigma} \left(
\pi{\lambda} e^{b \phi} + {Q \phi K } \right)~,
\label{liouville}
\ee
by analytic continuation $X^0\to i \phi$, $\beta \to
-ib$ and $V_0 \to -iQ$. This theory  was studied in
\cite{Fateev:2000ik,Teschner:2000md} and the analytic
continuation to (\ref{vko})
was studied in \cite{Gutperle:2003xf}.

In the
presence of the source, the free closed string field equations 
can be written as
 \be \label{eom}
 (Q + \bar Q) |\Psi \rangle = |B\rangle
 \ee
where the boundary state $|B\rangle $ obeys
 \be \label{consv}
(Q + \bar Q) |B\rangle = 0.
 \ee
$Q$ here is the BRST
charge in the linear dilaton background with the Virasoro
operators
 \be
L_m =
\half \sum_{n} : \a_{m-n}^\mu \a_{n\mu} :
~+~ \frac{i}{\sqrt{2}} (m+1) V_\nu\a^\nu_m~.
 \ee
$|\Psi \rangle$ and $|B\rangle $ contain the relevant terms:
 \bear
  |B \rangle & \supset &
 \left[\tilde A_{\mu \nu} (p) \a_{-1}^\mu \bar
 \a_{-1}^\nu - \tilde B (p)
 ( \bar b_{-1}
 c_{-1} + b_{-1} \bar c_{-1}) \right] \, c_0^+ \, c_1 \bar c_1  |p\rangle
 \\
 |\Psi \rangle &\supset  & \left[\tilde h_{\mu \nu} (p) \a_{-1}^\mu \bar
 \a_{-1}^\nu - \tilde \phi (p)
 ( \bar b_{-1}
 c_{-1} + b_{-1} \bar c_{-1}) \right] \, c_1 \bar c_1  |p\rangle~,
 \eear
where $\sqrt{2} \alpha_0^\mu |p \rangle = p^\mu |p \rangle $.
The equations of motion (\ref{eom}) and (\ref{consv}) lead to
  \bear \label{eqa}
\half p \cdot( p + 2i V)  \tilde h_{\mu \nu} (p) &=& \tilde  A_{\mu \nu} (p) \\
 \label{eqb}
 \half p \cdot( p + 2i V) \tilde \phi (p) &=&  \tilde B (p) \\
 (p^\mu + 2 i V^\mu)  \tilde A_{\mu \nu}(p) - p_\nu \tilde B (p)
 &=& 0 \label{mocon}
 \eear
under the gauge condition
\be
(p^\mu + 2 i V^\mu) \tilde h_{\mu\nu}(p) = p_\nu\tilde\phi(p)~.
\ee
To move to position space, we define
\be
f(x) \equiv \int \frac{d^{D}k}{(2 \pi)^D} ~e^{i p_\mu x^\mu} ~f(p)~,
\ee
for $f \in \{\tilde A_{\mu\nu}, \tilde B, \tilde h_{\mu\nu},
\tilde \phi\}$, and with $p_\mu = k_\mu - iV_\mu$ where $k_\mu$
is real. $\tilde A_{\mu\nu}$, $\tilde B$, $\tilde h_{\mu\nu}$, $\tilde \phi$
$(p=k-iV)$ have support as $k_\mu$ varies continuously in the $D-p$
directions transverse to the decaying brane (including the time direction),
but only for $k_\mu=0$ in the parallel directions.
Equations (\ref{eqa})-(\ref{mocon}) then become
 \bear
\label{eq1}
\half (-\partial^2 + 2 V^\mu\partial_\mu)\tilde h_{\mu\nu}(x)
&=& \tilde A_{\mu\nu}(x) \\
\label{eq2}
\half (-\partial^2 +2 V^\mu\partial_\mu )\tilde \phi(x) &=& \tilde B(x) \\
\label{conservation}
 (\partial^\mu - 2V^\mu)  \tilde A_{\mu \nu}(x) &=&
\partial_\nu \tilde B(x)
 \eear
and the gauge condition is
\be
 (\partial^\mu - 2V^\mu) \tilde h_{\mu \nu}(x)=  \partial_\nu
\tilde \phi(x)~.
\label{gauge}
 \ee

We need to relate $\tilde A_{\mu \nu}$ and $\tilde B$  to the
spacetime stress-energy tensor.
 Consider the string frame effective action
 \be
 S = \int d^D x \, \sqrt{-g} e^{- 2 \Phi} \left[4 V^2 + R + 4
 (\nabla \Phi)^2 \right] + S_{matter}~.
 \ee
Varying this action with respect to the string frame metric
$g_{\mu\nu}$, we obtain
\be
e^{-2\Phi} \left ( G_{\mu\nu} -2g_{\mu\nu} \nabla^2\Phi+
2 \nabla_\mu\nabla_\nu \Phi + 2g_{\mu\nu}(\nabla\Phi)^2
-2 V^2 g_{\mu\nu}
\right ) = T_{\mu\nu}~,
\label{fstring}
\ee
where $T_{\mu\nu}$ is the string frame stress tensor,
defined as the variation of $S_{matter}$ w.r.t $g^{\mu\nu}$.
Varying with respect to the dilaton, we obtain
\be
e^{-2\Phi} \left (2 R +8 V^2 - 8(\nabla\Phi)^2
+8\nabla^2\Phi
\right ) = \frac{\delta S_{matter}}{\delta \Phi} \equiv U~.
\label{fdila}
\ee
Expanding  (\ref{fstring})--(\ref{fdila}) around the
linear dilaton background
\be
 g_{\mu \nu} = \eta_{\mu \nu} + h_{\mu \nu}~,
~ \Phi = V_\mu x^\mu +  \phi~,
\ee
in the gauge
\be \label{ligauge}
(\partial^\mu-2 V^\mu) h_{\mu \nu} = \partial_\nu
(\half h - 2 \phi)~,
\ee
one finds that at the linear level
 \bear
 \label{anone}
- \half \partial^2 h_{\mu \nu} + V_\lam \partial^\lam h_{\mu \nu} &=&
 e^{2 V \cdot x} (T_{\mu \nu}+\frac{1}{4}\eta_{\mu\nu} U) \\
 \label{antwo}
\left (- \half \partial^2  + V^\mu \partial_\mu
\right )\left(\half h - 2 \phi\right)
&=& \frac{1}{4}e^{2 V \cdot x}U~.
 \eear
Comparing the above two equations to (\ref{eq1})--(\ref{eq2}), we find
\bear
h_{\mu\nu} = \tilde h_{\mu\nu}&~\quad~&
\half h - 2\phi = \tilde \phi \\
T_{\mu\nu} = e^{-2V \cdot x} (\tilde A_{\mu\nu} -
\tilde B \eta_{\mu\nu})
&~\quad~&
\frac{1}{4}U =  e^{-2 V \cdot x} \tilde B
\eear
Under these identifications, the gauge conditions
(\ref{ligauge}) and (\ref{gauge}) are one and the same.
The conservation equation (\ref{conservation})
gives a modified conservation law for the
stress tensor
\be
\partial^{\mu}T_{\mu\nu} =
\half V_\nu U~.
\label{modified.conservation}
\ee
The source term is due to the linear dilaton
breaking spacetime translational symmetry.
This equation is the same as equation
(\ref{modified.conservation.intro}), obtained
in section \ref{sec:ld}.

The boundary state coefficient
$\tilde B (x)$ is given by a disc worldsheet one point function
\footnote{
We note that the correlators on the right hand side contain a factor of
$e^{-V \cdot X}$ arising from the worldsheet action (\ref{vko}).}
 \bear
 \label{wsb}
\tilde B (x)~&=&   e^{V \cdot x}\int \frac{d^{D}k}{(2\pi)^D}
\, e^{i k \cdot x }\langle  e^{-i k \cdot X
+ V \cdot X}  \rangle ~.
\eear
(\ref{wsb}) was computed in \cite{Gutperle:2003xf}
by analytic continuation from the results of  \cite{Fateev:2000ik,
Teschner:2000md}:
\be
e^{-V\cdot x} \tilde B(x^0) =
 \frac{1}{\beta} \int \frac{d^{D-p} k}{(2\pi)^{D-p}}
e^{ik \cdot x} \tilde \lambda^{i k_0 /\beta}
{\Gamma(-ik_0/\beta)}{\Gamma(1+i\beta k_0)}~,
\ee
where
\be
\tilde \lam \equiv \pi\lam/\Gamma(1+\b^2)~.
\ee
Adopting the convention that $t=x^0$,
we derive an explicit expression for $\tilde B$ as follows:
\bear
e^{-V\cdot x} \tilde B(t) &=&
\frac{1}{\beta} \int \frac{d k_0}{2\pi} e^{i k_0 t}
\tilde \lambda^{i k_0 /\beta}
\int dq~ e^{-q}~ q^{-i k_0/\beta-1}
\int ds~ e^{-s}~ s^{i k_0 \beta}
\nn \\ &=&
\int \frac{d k_0}{2\pi} e^{i k_0 t}
\tilde \lambda^{i k_0 /\beta}
\int dt'~ e^{-e^{\b t'}}~ e^{-i k_0 t'}
\int ds~ e^{-s}~ s^{i k_0 \beta}
\nn \\ &=&
\int ds \exp\left(-s-\tilde \lambda e^{\b t} s^{\b^2}\right)~,
\eear
where we have suppressed the transverse delta functions
projecting onto the brane.
$\tilde A_{\mu\nu}$ is a one point function of a
descendant of the primary in (\ref{wsb}).  It must
be diagonal, since the total CFT describing the
string is a direct sum of the CFTs describing each
coordinate, and a left--right level matching condition must be
satisfied for each spacelike CFT separately for a nonzero
one point function.
A simple way to compute $\tilde A_{\mu\nu}$ is
to use the conservation law (\ref{mocon}).
Starting with the $00$ component, $\tilde A_{00}$:
\be
\tilde A_{00}(p) = -\frac{k_0 -i V_0}{k_0 +i V_0} \tilde B(p)~,
\ee
we obtain
\be
e^{-V\cdot x} \tilde A_{00}(t) =
 \frac{1}{\beta} \int \frac{d k_0}{(2\pi)}
e^{ik_0 t}  \frac{-k_0 +i V_0}{k_0 +i V_0}
\tilde \lambda^{i k_0 /\beta}
{\Gamma(-ik_0/\beta)}{\Gamma(1+i\beta k_0)}~.
\ee
Write
\be
 \frac{-k_0 +i V_0}{k_0 +i V_0} =
2 \left ( \int_0^1 dy ~y^{i k_0 /V_0}\right) - 1~.
\ee
Then, through a derivation similar to that for
$\tilde B$,
\bear
&&e^{-V\cdot x} \tilde A_{00}(t) = \\ \nn
&&\int ds \left [
2 \int_0^1 dy~ \exp\left(-s-\tilde \lambda e^{\b t} s^{\b^2} y^
{-\b/V_0}\right)
-\exp\left(-s-\tilde \lambda e^{\b t} s^{\b^2}\right)
\right ]~.
\eear
In addition, one easily finds that $\tilde A_{ii} = -\tilde B$.

Since $T_{ab}$ is zero for transverse directions $a,b$,
we focus below on the longitudinal directions $i,j$.
An explicit expression for $T_{\mu \nu}$ follows from above, 
again suppressing the transverse delta function
projecting onto the brane:
 \bear
T_{00}(x) &=& e^{-2V\cdot x}(\tilde A_{00} +\tilde B) \nn\\&=&
2 e^{-V\cdot x} \int_0^1  dy \int_0^\infty ds \,
 \exp\left ( -s - \tilde \lambda e^{\beta t } s^{\beta^2} y^\gamma
\right )
 \eear
and
 \be
 T_{ij} = -2 e^{-V \cdot x}\delta_{ij}
 \int_0^\infty ds \, \exp \left(- s - \tilde \lambda e^{\beta t} \,
 s^{\beta^2}\right)~.
 \ee
Also, we obtain the dilaton source term
\be
U = 4 e^{-V \cdot x} \int_0^\infty ds \,
\exp \left(- s - \tilde \lambda e^{\beta t} \,  s^{\beta^2}\right)~.
\ee
In the far past, before the brane decays, we find that
\bear
T_{00} (t \rightarrow -\infty) &=& 2 e^{-V \cdot x}\\
T_{ij} (t \rightarrow -\infty) &=& -2 e^{-V \cdot x} \delta_{ij}~.
\eear
In the far future, after the brane has decayed,
\bear
T_{00}(t \rightarrow \infty) & \sim&  e^{-V_i x^i}\\
T_{ij}(t \rightarrow \infty) &\sim& e^{- { \beta} t - V_i x^i} \delta_{ij}~.
\eear
Note that for $t \to -\infty$ the components of the stress tensor of
the D-brane scale as ${1 \over g_s (t)}$.  This is expected before
the tachyon condenses. For $t \to \infty$, $T_{00}$
becomes {\bf independent} of time, while $T_{ij}$
decays to zero. Thus we have tachyon dust in the far future. The
fact that, despite the time dependence of the string coupling, the
energy density becomes constant and the pressure goes to zero
corroborates the suggestion \cite{Lambert:2003zr,
Sen:2003bc, Shiraz} that this tachyon dust is a gas of
massive closed strings, whose energy is not affected by
a change in the dilaton the way the energy of a D-brane is.
The spatial dependence of the energy density for $t\rightarrow \infty$
can be explained from the assumption that the initial
D-brane energy is converted to the tachyon dust during a
time period around $x^0=0$. This initial energy profile is
imprinted in the tachyon dust and subsequently unchanged.

Note that the dust energy is still proportional to $1/g_s^0$ with
$g_s^0$ the constant part of string coupling. This implies that
the (order $g_s^0$) energy in closed string emission computed
from the annulus 
in the next section is not sufficient to account for the energy
in the dust. So, if the tachyon dust is indeed made of massive closed
strings, in this context their origin can not be understood from the
annulus.

\begin{figure}[t]
\centerline{\epsfbox{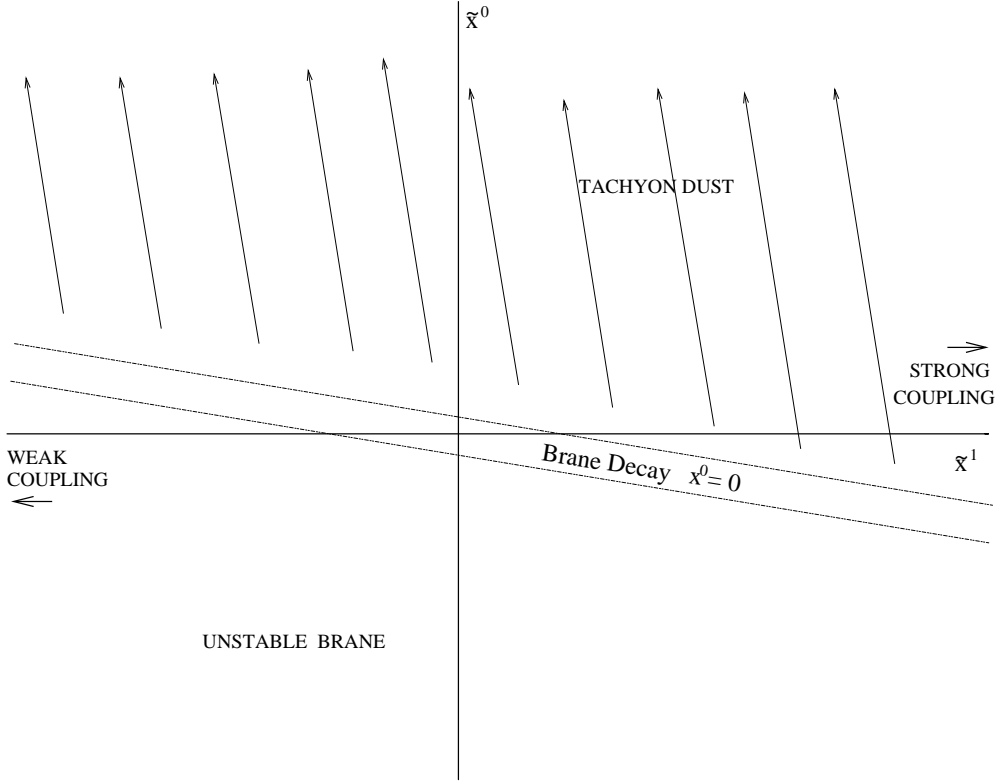}}
\caption{Brane decay in the frame in which the dilaton is time-independent
but grows with $\tilde x^1$. In this frame, the tachyon dust produced by a
normalizable decay mode drifts towards weak coupling with constant
velocity. Both the energy density of the dust and the energy density
of the brane which produced it decrease exponentially towards the
strong coupling region.}
\label{figure2}
\end{figure}

The Lorentz frame we have discussed so far is a natural one
because the brane decay occurs simultaneously in all space.
In the following sections we shall also see that it is the
simplest frame for CFT calculations. However, for a spacelike dilaton
$V_\mu V^\mu \equiv |V|^2 >0$ there
is a timelike Killing symmetry which is obscured in this frame.
For simplicity let us assume that the dilaton gradient points in the
$x^1$ direction so that
\be
\Phi(x)=V_0 x^0+V_1x^1~~~~~T(x)=e^{\beta x^0},
\label{frv}
\ee
with $V_1 > 0$.
Then the dilaton background has
a timelike Killing vector
\be
K=\frac{1}{|V|}(V^1\partial_0+V^0\partial_1)~.
\ee
This allows for the construction of a conserved
energy-momentum four-vector
\be
j_\nu=K^\mu T_{\mu\nu}~,\qquad\partial_\mu
j^\mu=0~.
\ee
Since $T_{0i}=0$ in the frame (\ref{frv}), the tachyon
dust has no momentum there. Clearly if
$V_0 > 0$, the dust is moving towards strong coupling;
if $V_0 < 0$, it is moving towards weak coupling.

Of particular interest is the case of a normalizable tachyon, in
the sense discussed in section \ref{sec:ld}, as the physical decay of
the brane would presumably proceed via such normalizable modes.
It is natural to consider this issue in the Lorentz frame
\be
\Phi(\tilde x)=|V|\tilde x^1~~~~~T(\tilde x)=
\exp\left(\frac{\beta (V_1\tilde x^0 -V_0\tilde x^1)}{|V|}\right),
\label{frvs}
\ee
in which the
dilaton is purely spatial and energy is conserved.
For the tachyon to be normalizable in this frame
we must have
\be
2 \beta V_0 = - V^2 < 0~.
\ee
If we consider a spatially oscillating profile, $\beta$ becomes
smaller but the sign restriction on $V_0$ remains.
Thus, if we demand normalizability, the tachyon dust slides
towards weak coupling.  At any given fixed spatial
point (i.e. fixed value of the string coupling)
in this frame, the energy decreases exponentially with
time, and the dust disappears. This is illustrated in Figure \ref{figure2}.


\section{Closed string emission}
\label{sec:closed}

In this section, we will define and compute the annulus diagram, extract its
imaginary part, compare with the closed string emission
computed in  \cite{Lambert:2003zr} and discuss the various
divergences.
As in section \ref{sec:stress}, we boost into a frame where the
tachyon profile is $\exp(\beta X^0)$, and take an arbitrary linear
dilaton with $V_0 = \beta - 1/\beta$.

\subsection{Partition function}
\label{subsec:partition}

We would like to compute the open string partition function,
which in the closed string channel is written as
\be
Z(\tilde q) = \langle B|\tilde q^{L_0 -\frac{D-2}{24}}|B\rangle~.
\ee
This expression factorizes into the time-like and the space-like part.
We will first study the time-like part in detail.  It can be written as
\cite{Teschner:2000md}
\be
Z^{X^0}(\tilde q)=
\int \frac{d \omega}{2 \pi} |U(\alpha)|^2 \chi_\alpha(\tilde q)~,
\label{rts}
\ee
where $U(\alpha)$ is the one-point function of $e^{-2 i \alpha X^0}$
on the disc.
The operator $e^{-2i\alpha X^0}$ must be dressed,
so $2 \alpha = \omega + i V_0$, and the integration is
over all real frequencies $\omega$. This amounts to
including only oscillating closed strings
which behave as $e^{i\omega X^0}$ rather than $e^{\omega X^0}$.
$\chi_\alpha$ is the character,
\be
\chi_\alpha(q)=\eta^{-1}(q)~ q^{-(\alpha-iV_0/2)^2}~.
\ee
The bulk one point function for the
insertion of a dressed closed string vertex operator is
\cite{Fateev:2000ik,
Gutperle:2003xf}
\be
U(\alpha) \equiv
\langle e^{(V_0 - i \omega)X^0}\rangle =-
\frac{i}{\beta} \tilde \lambda^{\frac {i\omega}{\beta}} \Gamma(1+i\omega \beta) \Gamma(-i \omega/\beta)~,
\label{eqn:thermo:I(E)}
\ee
and we obtain
\be
|U(\alpha)|^2 =
~\frac{\pi^2}{ \sinh(\pi \omega /\beta) \sinh(\pi \omega \beta)}~.
\label{eqn:|U|}
\ee
(\ref{rts}) may be now written as
\be
Z^{X^0}(\tilde q) = \int_{-\infty}^\infty
\frac{d\omega}{2 } \frac{ \pi~ \chi_{(\omega +i V_0)/2}(\tilde q)}
{\sinh(\pi \beta \omega)\sinh(\pi \omega/\beta)}~.
\label{rsts}
\ee

Unfortunately (\ref{rsts}) is
ill-defined because the character goes as
$q^{-\omega^2}$ for large $\omega$ corresponding to the negative $L_0$
eigenvalues for timelike modes. Our strategy will be to formally
manipulate
(\ref{rsts}) into a form which is well-defined, which will then be adopted
as the definition of $Z^{X^0}$.

We begin with the observation that the ``oscillating''
closed string characters
can be expressed as a modular transform of non-oscillating open string
characters (which are never on-shell) :
\be
\chi_\alpha(\tilde q)=
\sqrt{2}\int_{-\infty} ^\infty d{\nu} \cosh(4\pi(\alpha-iV_0/2){\nu})
\chi_{i({\nu}+ V_0 /2)}(q)~,\ee  where
\be q=e^{-\frac{\pi^2}{t} },\qquad \tilde q=e^{-4 t}
\ee
are modular transforms of one another.
The RHS of (\ref{rsts}) then becomes
\be
\frac{\pi}{\sqrt{2}}
\int_{-\infty}^\infty
d\omega~\int_{-\infty} ^\infty d{\nu}
\left ( \frac{\cosh(2\pi\omega {\nu})} {\sinh(\pi
\beta \omega)\sinh(\pi \omega/\beta)} \right )
\chi_{i(\nu+V_0/2)}(q)~.
\ee
The expression above has a double pole at $\omega=0$, which can
be taken care of by a
subtraction\footnote{The same subtraction
was used in \cite{Teschner:2000md} to regulate the
partition function; it not change the value
of the integral, since a double pole has no residue.}
\be
\frac{\pi}{\sqrt{2}}
\int_{-\infty}^\infty
d\omega~\int_{-\infty} ^\infty d{\nu}
\left ( \frac{ \cosh(2\pi\omega {\nu})} {\sinh(\pi
\beta \omega)\sinh(\pi \omega/\beta)}
  - \frac{1}{\pi^2\omega^2}\right )
\chi_{i(\nu+V_0/2)}(q)~. \label{yss}
\ee
A convergent expression for $Z^{X^0}$ can now be found
using the integral representation of the $q$-gamma function
\cite{Fateev:2000ik}
\be
\partial_x \ln S_\beta(x)=-\int_{-\infty}^\infty dt
\left(
\frac{\cosh((2x-Q_\beta)t)}
{2 \sinh \beta t \sinh \frac{t}{\beta}}
- \frac{1}{2t^2}
\right)
\ee
(where ${Q_\beta}\equiv \beta+\frac{1}{\beta}$)
and interchanging the integration order in (\ref{yss}).
This gives our
definition of $Z^{X^0}$ in terms of the convergent integral
\be
Z^{X^0}(q)\equiv\sqrt 2
\int_{-\infty}^\infty d\nu \chi_{i(\nu +V_0/2)}(q)
\partial_\nu \ln S_\beta\left(\frac{Q_\beta}{2}
-\nu  \right)~.
\label{zdef}
\ee
The function
$S_\beta(\frac{Q_\beta}{2} -{\nu}     ) $ has simple
poles at $\nu =(m+\half)\beta +\frac{n+\half}{\beta}$ for non-negative
integer $m,n$
and simple zeroes for negative integer $m,n$. We take the contour in
(\ref{zdef}) to go just above the real axis for positive $\nu$ and just
below
for negative $\nu$.

$Z^{X^0}$ acquires an imaginary part from integrating in the $\nu$ plane
past the poles
\be
 \Im Z^{X^0}(q)
= 2 \sqrt{2} \pi~ \eta^{-1}(q)\sum_{m,n=0}^\infty
e^{-((n+\half)\beta +(m+\half)/\beta)^2\frac{\pi^2}{t}}~.
 \ee

The space-like part (including ghosts) of the partition function is standard
\bear
Z^{X^i+bc}(q) &=& V_p\int \frac{d^{D-1-p}k_\perp}{(2\pi)^{D-1-p}}
~\tilde q^{k^2_\perp/4}~
\eta^{3-D}(\tilde q) \nn \\&=&
\frac{{\cal N}^{-2}_p V_p}{2\sqrt 2\pi t}
\int\frac{d^{p}k_\|}{(2\pi)^p}~ q^{k^2_\|}~
\eta^{3-D}(q)~,
\eear
where ${\cal N}_p = \pi^{(D-4)/4} (2\pi)^{(D-2)/4 - p}$ is the
normalization of the boundary state, same as for the non-decaying brane.
Putting the two things together, we obtain
\bear
 Z(q) &=& {\cal N}_p^2 Z^{X^i+bc}Z^{X^0} \nn\\ &=& \frac{V_p}{2 \pi t}
\int_{-\infty}^\infty d\nu \int \frac{d^{p}k_\|}{(2\pi)^p}
 ~q^{\nu^2+k_{\|}^2 }~\eta^{2-D}(q)~
\partial_\nu \ln S_\beta\left(
\frac{Q_\beta}{2}-\nu \right).
\eear
The imaginary part from the poles is
\be
 \Im Z(q)
= \frac{V_p}{t} \int \frac{d^{p}k_\|}{(2\pi)^p} ~q^{k_{\|}^2 }~
 \eta^{2-D}(q)\sum_{m,n=0}^\infty
e^{-((n+\half)\beta +(m+\half)/\beta)^2\frac{\pi^2}{t}}
~.
\label{img}
\ee
Without the linear dilaton,  $\beta=1, V_\mu=0$, and the above expression
simplifies to
\be
 \Im Z(q)
=\frac{V_p}{t} \int \frac{d^{p}k_\|}{(2\pi)^p}~ q^{k_{\|}^2 }~
\eta^{-24}(\tilde q)\sum_{n=1}^\infty n
e^{-n^2\frac{\pi^2}{t}}~.
\label{img0}
 \ee
In subsection \ref{subsec:closed}, we shall see that this agrees with
the expected answer for closed string emission.
But first, we want to consider whether the annulus diagram
might have any other contributions to the imaginary part.

\subsection{Annulus diagram}
\label{subsec:annulus}

In principle the full imaginary part of the annulus
could contain additional terms associated with divergences in
the integration of the partition function $Z(q)$ over the modulus.
In the computation of, for example,
 open string pair production in an  electric field,
such imaginary parts arise near poles from on-shell open string states.
We will not find such imaginary parts here but the form of the  integrated
partition function is nevertheless illuminating.

Writing $q=e^{-s}$ ($s=\pi^2/t$),
adding an integration over $s$ and writing the $\eta$ functions
as a sum over open string oscillators $N$ yields
\bear
{\bf Z}&=&\int {ds}~ Z\left(\frac{is}{2\pi}\right)\\&=&\nn
V_p \int \frac{ds}{2 \pi s} \int_{-\infty}^\infty d\nu \int
\frac{d^pk_{\|}}{(2\pi)^p} \sum_N
e^{-s(\nu^2+k_{\|}^2+N-1) }
\partial_\nu \ln S_\beta\left(
\frac{Q_\beta}{2}-\nu  \right)~.
\eear
We will evaluate this by defining ${\bf Z} \equiv V_p \int
\frac{d^pk_{\|}}{(2\pi)^p}
{\bf Z}( k_{\|}^2)$ so that
\bear
- \frac{d}{dk_{\|}^2} {\bf Z}(
k_{\|}^2)&=& \int \frac{ds}{2\pi} \int_{-\infty}^\infty d\nu  \sum_N
e^{-s(\nu^2+k_{\|}^2+N-1) }
\partial_\nu \ln S_\beta\left(
\frac{Q_\beta}{2}-\nu  \right)
\nn\\ \\ \nn
&=& \frac{1}{2\pi} \int_{-\infty}^\infty d\nu  \sum_N
\frac{1}{\nu^2+k_{\|}^2+N-1 }
\partial_\nu \ln S_\beta\left(
\frac{Q_\beta}{2}-\nu  \right)  .
\eear
The $\nu$ integration can now be done by closing the contour in the
upper half-plane. The result picks up poles from
$\partial S_\beta$, as well as a contribution from the
pole at $\nu=i\sqrt{k_{\|}^2+N-1} $ corresponding to on-shell
open strings. One finds
\bear
-\frac{d}{dk_{\|}^2} {\bf Z}(
k_{\|}^2)&=&\sum_N \sum_{m,n=0}^\infty \frac{i }{((n+\half)\beta
+(m+\half)/\beta)^2+k_{\|}^2+N-1 }
\nn\\&+&\sum_N
\frac{i }{2 \nu }
\partial_{\nu} \ln S_\beta\left(
\frac{Q_\beta}{2}- \nu \right ) |_{ \nu=i\sqrt{k_{\|}^2+N-1} }
-\frac{\pi}{2}~,
\eear
where the last term comes from subtracting the contour at infinity.
Integrating with respect to $k_{\|}^2$ gives
\bear
{\bf Z}(
k_{\|}^2)&=&V_p \sum_N i \left[\sum_{m,n=0}^\infty \right .
- \ln \left ( ((n+\half)\beta
+(m+\half)/\beta)^2+k_{\|}^2+N-1 ) \right )
\nn \\  &+&  \left . \ln S_\beta\left(
\frac{Q_\beta}{2}- i\sqrt{k_{\|}^2+N-1}\right)
-\frac{i\pi k_{\|}^2}{2} +
\mbox{constant} \right]~,
\eear
where the leading large $k_{\|}^2 $ behavior of $S_\beta$ just cancels the
$\frac{i\pi k_{\|}^2}{2}$  term.
${\bf Z}$ is then
\bear
{\bf Z}&=& \sum_N V_p \int \frac{d^pk_{\|}}{(2\pi)^p}
\left[-i \sum_{m,n=0}^\infty\ln [(n+\half)\beta
+(m+\half)/\beta)^2+k_{\|}^2+N-1 ] \right .
\nn \\ &+&\left .i \ln S_\beta\left(
\frac{Q_\beta}{2}-i\sqrt{k_{\|}^2+N-1} \right )+\frac{\pi k_{\|}^2}{2}  +
\mbox{constant}  \right].
\label{ccb}
\eear
The first part reproduces (\ref{img}).  Using the identity
\cite{Fateev:2000ik}
$S_\beta(x)S_\beta(Q_\beta-x)=1$,
 we see that the second term, which arises at the pole for on-shell
open string states,  has no imaginary
part. This suggests that the imaginary part does not contain
an extra contribution from open string pair production.

Divergences remain in (\ref{ccb}) from both closed and open string
tachyons. Presumably the divergence associated with the closed string
tachyon
will be absent in the superstring. But even there, an unstable brane
always has an open string tachyon. The divergence arises because
there is no vacuum state for the incoming tachyon. Quantum
fluctuations prevent one from perching the tachyon at the maximum
of its potential all the way to the infinite past. Therefore initial
conditions must be imposed at some finite time, which complicates
the analysis. A more physical context to study this is in the
full s-brane, in which the tachyon remains only momentarily at its
maximum.  For the full s-brane in the superstring we do not expect any
divergences from tachyons, if the initial conditions are chosen
correctly.

\subsection{Closed string emission and its divergences}
\label{subsec:closed}

Performing a modular transform on (\ref{img}), and integrating
over $s=\pi^2/t$, we can write it in the form
\be
\Im {\bf Z} = V_p\sum_{n,m=0}^{\infty}
\int_0^{\infty} \frac {ds}{s}
\frac{1}{(4\pi s)^{\half{p}}} ~
e^{-\left((n+\half) \beta +(m+\half)/ \beta \right)^2 s}~
\frac{1}{\eta^{D-2} \left ( \frac{is}{2\pi} \right ) }~.
\label{eqn:Z}
\ee
For the special case, $V_\mu=0, \beta=1$, this simplifies
to
\be
\Im {\bf Z} = V_p\sum_{n=1}^{\infty} n \int_0^{\infty} \frac {ds}{s}
\frac{1}{(4\pi s)^{\half{p}}} ~e ^{-n^2 s}~ \frac{1}{\eta^{24}
\left ( \frac{is}{2\pi} \right ) }~.
\label{eqn:Z0}
\ee
Closed string emission (without the linear dilaton) was computed in
\cite{Lambert:2003zr}.  In the appendix, we review that computation
and extend it to the linear dilaton background.  The answers
we obtain agree precisely with the above expressions for
$\Im {\bf Z}$.
Thus, we can interpret the imaginary part of the annulus diagram
as the amplitude for closed string emission.

We are interested in its divergences, begining our analysis with
the $V_\mu=0$ case.
The expression (\ref{eqn:Z0})
above has potential for both open string IR
divergences (large s) and UV divergences (small s).

For large $s$, we have that $\eta(i s/2\pi) \sim \exp(-s / 24)$
plus exponentially suppressed corrections, thus, in the IR, the
expression is divergent if $p = 0$ and convergent for larger $p$.
This corresponds to a UV divergence in the emission of highly
massive closed strings. It can be understood as originating from
massless open strings stretching between two sD-branes
\cite{Lambert:2003zr}.

Similar expressions exist for the expectation value of powers of
energy in the closed string emission. These expressions contain an
extra power of $s$ for every power of energy. Thus, the
expectation value of $E^m$ will be divergent for $p \leq 2m$. This
indicates that the system becomes strongly coupled for any $p$.

In the UV (at small s) the expression is approximately
\be
V_p
\int_0
\frac {ds}{2 s^2} \frac{1}{(4\pi s)^{\half{p}}}~ \left (
\frac{s}{2\pi} \right )^{12} \left ( e^{\frac{4\pi^2}{s}} + 24 +
\ldots \right )~. \ee After we throw out the highly divergent first
term (which is due to the closed string tachyon), there is still a
power-law UV divergence for $p \ge 22$.

These divergences signal the breakdown of string perturbation
theory. Let us now examine the expression including the
linear dilaton, (\ref{eqn:Z}).

The UV region of the closed string emission is
$E \to \infty$, or equivalently $s \to \infty$ (open string IR).
In that limit, we obtain
\be
\Im {\bf Z} \sim
V_p \int^{\infty} \frac {ds}{s}
\frac{1}{(4\pi s)^{\half{p}}} ~
e^{-\frac{1}{4}\left(\beta +1/ \beta \right)^2 s}~
e^{\frac{D-2}{24}s}~.
\label{frgb}
\ee
The exponent is negative as long as $V_i\neq  0$, since
\be
-\frac{1}{4}\left ( (\b + 1/\b)\right)^2 + \frac{D-2}{24}
= -\frac {V_i^2}{4}~,
\ee
where $V_i$ is the spacelike part of the dilaton gradient.
There is no UV divergence as long as the dilaton is not precisely
aligned with the timelike direction of the decay. This is of course always
the case if the dilaton gradient is spacelike.
Since we would like to make a Lorentz invariant statement 
about finite production for
all exponential decay modes of the tachyon $e^{t_\mu X^\mu}$,
we shall restrict our claims to spacelike dilaton gradients.

Expression (\ref{eqn:Z}) can be interpreted as the partition function
for open strings stretched between an array of sD-branes at imaginary time.
These are located at $X^0 = \sigma (n+\half) \beta$ for $n \geq 0$ and at
$X^0 = -\sigma(m+\half) / \beta$ for $m \geq 0$, with
$\sigma = \pm i$.  A UV divergence in closed string emission transforms,
under a modular transformation, into an IR divergence arising from a
massless open string.  Here, the mass of the lightest open
string state is
\be
m_{\rm lightest}^2=\left ( \half (\b + 1/\b)\right)^2 - \frac{D-2}{24}
= \frac {V_i^2}{4}~,
\ee
in agreement with above discussion.

There are also potential UV divergences in the open string channel.
These appear in the expansion at
small s:
\be V_p
\int_0
\frac {ds}{2 s^2}
\frac{1}{(4\pi s)^{\half{p}}}~
\left ( \frac{s}{2\pi} \right )^{\frac{D-2}{2}}
\left ( e^{\frac{(D-2)\pi^2}{6s}} +
(D-2)e^{\frac{(D-26)\pi^2}{6s}} + \ldots
\right )~.
\label{smalls}
\ee
Hence, for $D<26$, a spacelike dilaton, the UV divergence
(ignoring the first term, due to the closed string tachyon) is gone as well.

\subsection{Open-closed string duality}

For simplicity, in this subsection the linear dilaton
is set to zero ($\beta=1$).

\begin{figure}[t]
\centerline{\epsfbox{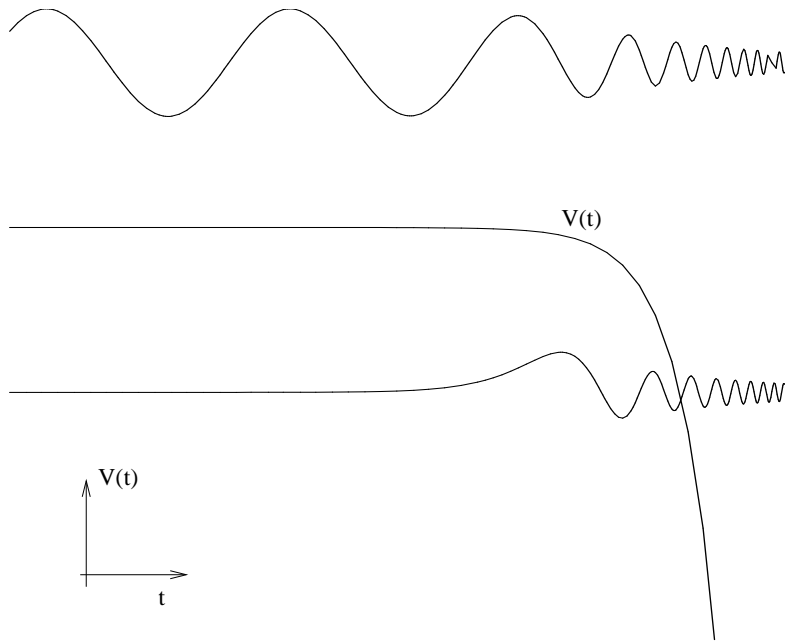}}
\caption{
In the unbounded from below potential V(t), there are two kinds
of eigenstates.  Those with positive total energy oscillate
in the far past, and correspond to on-shell strings propagating
on the brane.  Those with negative energy decay exponentially
in the far past and are always off-shell.  Poles in the 
spectral density of these unphysical states correspond
to closed string emission.}
\label{figure3}
\end{figure}

In subsection \ref{subsec:partition}, we wrote the partition
function as an integral over non-oscillating open string
states
\be
Z^{X^0}(q)\equiv
\int_{-\infty}^\infty d\nu
~\sqrt 2 \pi \nu \cot(\pi \nu)
~\chi_{i\nu}(q)
\label{Z.beta=1}
~.
\ee
In general, the partition function can be written as
\be
Z^{X^0}(q)\equiv
\int_{-\infty}^\infty d\nu
~N(\alpha)
~\chi_{\alpha}(q)
~,
\label{Z.open}
\ee
where $N(\a)$ the spectral density of open strings.
$N(\a)$ is not known, but (just as for ordinary Liouville theory)
we might be able to gain some insight into it by considering
the mini-superspace approximation.

In the mini-superspace approximation, the annulus diagram
can be written as
\be
{\bf Z} = V_p \,  \int_0^\infty {d s \over
  s} {1 \over \eta^{24} (is/2\pi)} {1 \over (4 \pi s)^{p \over 2}}
 ~\Tr { e^{s H}}~,
\label{minisuper}
 \ee
where $H$ is the Hamiltonian for a particle moving in one-dimension
 \be
 H = -\partial_t^2 - \lambda e^t~.
\ee
Notice that this looks real, but can develop imaginary
parts due to divergences.

The problem thus reduces to a
problem in quantum mechanics \cite{Strominger:2002pc,
Maloney:2003ck}
\be
-\partial_t^2 \phi(t) + V(t)\phi(t) = \omega^2\phi(t),\qquad
V(t) =  -\lambda e^t~,
\ee
where $t$ is the zero mode of $X^0$.
The solutions to the above equation are Bessel functions,
$\phi(t) \sim J_{2i\omega}(2\sqrt{\lambda}e^{t/2})$.  $\omega$
can either be real, or equal to $-i\nu$ with $\nu$ real and positive.

In the first branch (real $\omega$), the integral over
$s$ in equation (\ref{minisuper}) will diverge when the mode
under consideration is on-shell.  Analytic continuation controlling
this divergence will give $\bf Z$ an imaginary part, corresponding to
open string pair production.

In the second branch, open strings cannot go on shell.  However,
looking at equation (\ref{Z.beta=1}) and comparing
to (\ref{Z.open}), we realize that the spectral density
on this branch has poles as a function of $\nu$.  These poles
correspond to production of on-shell closed strings,
as computed in subsection \ref{subsec:partition}.

It is not clear that in a given boundary CFT (determined
by initial conditions for both closed and open strings) 
both these branches contribute to the annulus.


\section{Open string pair production}
\label{sec:open}

 In this section we compute the pair production of open strings
in a linear dilaton background along the lines discussed in
\cite{Strominger:2002pc,Gutperle:2003xf, Maloney:2003ck}.

\begin{figure}[t]
\centerline{\epsfbox{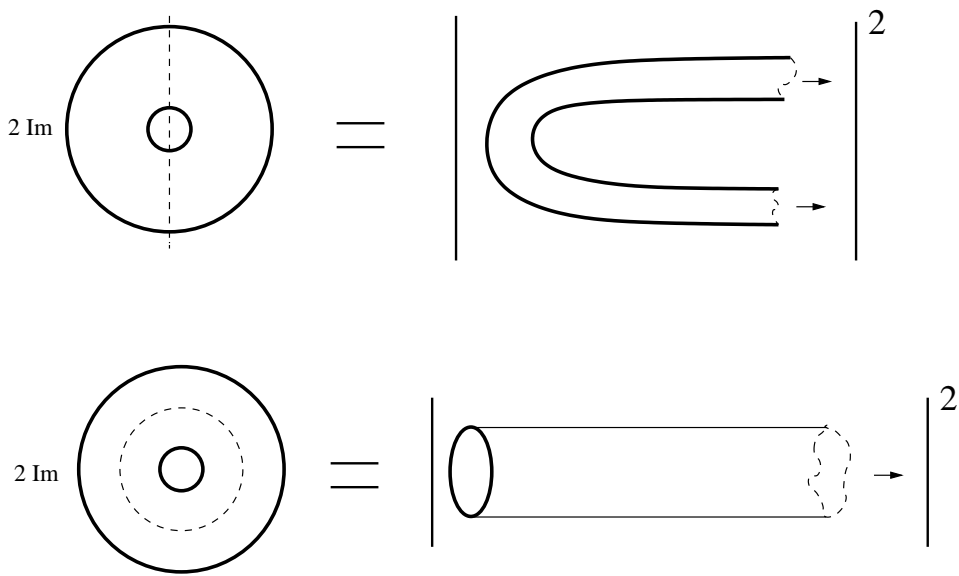}}
\caption{The annulus can be interpreted as the square of a closed string
one-point function or open string two point function, depending on how it
is sliced. Corresponding imaginary parts from on-shell
closed string emission or
open string pair production can appear in different kinematical regions of
the diagram.}
\label{figure1}
\end{figure}

Open strings are in general pair produced on a decaying brane because they
are governed by a time-dependent Hamiltonian. This implies that
the $in$ and $out$ vacua are not the same. The vacuum decay
amplitude is
\be
W= -{\rm Re}\ln \langle out | in \rangle~.
\label{ospp}
\ee
We work in an approximation in which the open strings
propagate freely on the brane so that (\ref{ospp}) is a one loop
computation. The interpretations of the annulus as either
open or closed string
production correspond to different slicings,
as illustrated in Figure \ref{figure1}.
(\ref{ospp}) is the
formula used by Schwinger to compute pair production in an
electric field. Note however that in the latter case there is an
additional volume divergence coming from the time direction,
whereas here the open string creation is localized in
time and there is no such divergence.

The $in$ and $out$ vacua are related by a Bogolubov transformation
\be
|out\rangle=\prod_{k_{\|}}(1-|\gamma_{k_{\|}}|^2)^{1/4}e^{-\half
\gamma_{k_{\|}}^*(a^{in\dagger}_{k_{\|}})^2 }|in\rangle~,
\ee
where  $\gamma_{k_{\|}}$ and $a^{in\dagger}_{k_{\|}}$ are
the Bogolubov coefficients and the creation operators associated to an
open string mode with longitudinal spatial momentum $k_{\|}$. (We
follow the notation of  \cite{Strominger:2002pc,Gutperle:2003xf,
Maloney:2003ck}.) It follows that
\be
\frac{W}{V_p}=-\frac{1}{4}\sum_N \int\frac{d^pk_{\|}}{(2\pi)^p}
\ln(1-|\gamma_{k_{\|}}|^2)~,
\label{cxz}
\ee
where
\be
\omega^2=k_{\|}^2+N- \frac{D-2}{24}~.\label{omg}
\ee

The Bogolubov coefficients $\gamma$ are conjecturally
related \cite{Gutperle:2003xf} to the
boundary two point function of dressed open string operators.
At large $\omega$ (and hence small $\gamma_\omega$):
\be
|\gamma_\omega|^2 \sim \left|d(\omega+\frac{iV_0 }{2})\right|^2~,
\ee
which then determines the convergence of (\ref{cxz}) at large $\omega$.
From formulae in \cite{Gutperle:2003xf, Fateev:2000ik},
the norm of the two point
function is\footnote{This follows from (4.7), (4.10) and (4.31) of
\cite{Gutperle:2003xf} upon substitution of $b=i\beta$, $Q=iV_0$.}
\be
\left|d (\omega+\frac{iV_0}{2})\right |^2
=\left|\frac{\beta S_{i\beta}(2\omega +iV_0)
\Gamma(1-\frac{2i\omega}{\beta})\Gamma(2i\beta\omega)}{2\pi S^2_{i\beta}(
\omega+iV_0/2)}\right|^2~.
\label{pkj}
\ee
$S_{i\beta}(x)$ is the ratio of Barnes double gamma functions employed in
\cite{Fateev:2000ik}.\footnote{In general $S_{i\beta}$ is ill defined for
real $\beta$ as discussed in \cite{Gutperle:2003xf} but the quantities we
compute here have a smooth limit as $\beta$ approaches the real axis.}
It has an expansion for large $x$ with $\pm {\rm Im}~x >0$ as
\be
\ln S_{i\beta}(x)= \pm \frac{i\pi}{2}\left[x^2-iV_0x +O(x^0)\right]~.
\label{sxp}
\ee
Similarly for large positive $x$ Stirling's formula gives
\be
{\rm Re}\ln \Gamma(ix)= -\frac{\pi x}{2} +O(x^0)~.
\ee
We wish to find the asymptotic behavior of (\ref{pkj}) for
large $\omega$.
If $\beta <1$, $V_0<0$ and we should use the lower sign in (\ref{sxp})
which yields
\be
\left|d(\omega+\frac{iV_0}{2})\right|^2=
e^{-\frac{4\pi \omega}{\beta}+O(\omega^0)}~.
\ee
On the other hand for $\beta>1$ we should use the upper sign which
yields instead
\be
\left|d(\omega+\frac{iV_0}{2})\right|^2=
e^{-{4\pi \omega\beta}+O(\omega^0)}~.
\ee
It follows that for large $\omega$ and $\beta<1$,
\be
\frac{W}{V_p} \sim \sum_N\int d^pk_{\|} e^{-\frac{4\pi \omega}{\beta}}
\sim \int d\omega e^{\frac{\omega}{T_H}-\frac{4\pi \omega}{\beta}}
\sim \int d\omega  e^{2\pi\omega \sqrt \frac{D-2}{{6}}-\frac{4\pi \omega}{\beta}},
\ee
where we have used expression (\ref{thag}) for $T_H$.
The corresponding expression for $\beta>1$ involves $\beta \to 1/\beta$.
This gives the production of highly massive open string pairs.
The exponent here is more negative than the one we
saw earlier in equation (\ref{frgb}) for the highly massive $closed$
strings. Hence the linear dilaton background
renders finite both the closed string emission and the open string
pair production rates.

Without the dilaton, $\b = 1$ and $D=26$, and the
two exponentials cancel.  There is then a power divergence for $p \ge 23$.


\section*{Acknowledgments}
We are grateful to A. Maloney, S. Minwalla, J. Polchinski,
A. Sen, T. Takayanagi, N. Toumbas
and Xi Yin for useful conversations. This work was supported in part by DOE
grant DE-FG02-91ER40654 (AS,JLK) and the Harvard Society of Fellows (JLK).
Work of HL is supported by DE-FG02-96ER40949.

\appendix

\section{Appendix One: Closed string emission}

In this appendix, we will compute closed string emission during
brane decay in various linear dilaton theories. We begin with a
review of \cite{Lambert:2003zr} for closed string emission during
brane decay (or half s-brane) without a dilaton\footnote{Nearly
identical results apply to the full s-brane with a
Hartle-Hawking-type contour prescription\cite{Lambert:2003zr}.}.
These formulae will be generalized to a linear dilaton background.

Flat space brane decay is described by a TBL string worldsheet CFT
with an action for the timelike coordinate $X^0$ \be -
\frac{1}{2\pi} \int_{\Sigma} \partial X^0
\partial X^0  +\frac{\lambda}{2}
\int_{\partial \Sigma}  e^{X^0} ~, \label{fcv} \ee where we have
set $\alpha'=1$. The one point function for this theory is
\cite{Fateev:2000ik,Gutperle:2003xf} 
\be 
U\left(\a = \frac{\omega}{2}\right) = \langle e^{-i \omega X^0} \rangle 
= {\tilde \lambda}^{i\omega} \frac{\pi}{\sinh \pi
\omega}~. \label{eqn:review:I(E)} \ee $ U(\a)  
$ is related to the number
of closed strings emitted by the half s-brane via
\cite{Lambert:2003zr}\be \frac {\bar N} {V_p} = {\cal{N}}_p^2
\sum_s \frac {1}{2 \omega_s} |U(\omega/2)|^2~, 
\label{eqn:review:N/Vdef} \ee
where ${\cal N}_p = \pi^{\frac{11}{2}} (2 \pi)^{6-p}$ is a
normalization constant. The sum is over all the closed strings
with identical left and right moving sectors, which is equivalent
to the sum over open strings. $\omega_s$ is the on-shell energy for a
given closed string state, \be \omega_s^2 = k^2_{\perp} + 4 N - 4~.
\label{eqn:review:onshell} \ee

It is useful to rewrite equation (\ref{eqn:review:N/Vdef}) in the
open string channel, where its divergences can be most easily
analyzed. Substitute equation (\ref{eqn:review:I(E)}) into
(\ref{eqn:review:N/Vdef}) and expand to obtain
\be \frac {\bar N}{V_p} =
{\cal {N}}_p^2 \sum_N \int
\frac{d^{25-p}k_{\perp}}{(2\pi)^{25-p}} \frac{1}{2 \omega}
\sum_{n=1}^{\infty} (4\pi n e^{-2 \pi \omega n})~. \ee The sum is over the
Hilbert space of a single open string and the integral is over all
momenta transverse to the brane.  $\omega$ is a function of $N$ and
$k_\perp$ as given by equation (\ref{eqn:review:onshell}). Now,
rewrite
\bear
\frac {1}{2\omega} e^{-2 \pi \omega n} &=& \frac{1}{2\pi}
\int d k_0 \frac {1}{k_0^2 +\omega^2} e^{2 \pi i k_0  n}
\\ \nn &=& \frac{1}{2\pi}
\int d k_0 \int_0^\infty ~dt
e^{-t(k_0^2 + k_{\perp}^2 + 4(N-1))} e^{2 \pi i k_0  n}~,
\eear
and perform the Gaussian integrals over $k_\perp$ and $k_0$ to obtain
\be \frac {\bar N} {V_p} =
C \sum_{n=1}^{\infty} n \int_o^\infty dt~
t^{-\frac{26-p}{2}} e^{-\frac{(\pi n)^2}{t}} \sum_N e^{-4 t (N-1)}
~. \ee The sum over all open string modes $N$ gives an eta
function and the answer, after a modular transform $s = (2 \pi)^2
/ (4 t)$, is \bear \frac {\bar N} {V_p} &=&
C \sum_{n=1}^{\infty} n
\int_0^\infty dt~ t^{-\frac{26-p}{2}} e^{-\frac{(\pi n)^2}{t}}
\eta^{-24} \left (\frac{4 ti}{2 \pi}  \right )
\nn \\
&=& \sum_{n=1}^{\infty} n \int_0^{\infty} \frac {ds}{s}
\frac{1}{(4\pi s)^{\half{p}}} ~e ^{-n^2 s}~ \frac{1}{\eta^{24}
\left ( \frac{is}{2\pi} \right ) }~. \label{eqn:review:N/V} \eear
$C$ in the above is a constant, $(2\pi)^{-12-p}\pi^{24-p/2}$.

We wish to evaluate formula
(\ref{eqn:review:N/Vdef}) for
$\frac{\bar N}{V_p}$ in the presence of a linear dilaton.
The squared one point function (\ref{eqn:thermo:I(E)}) is
\bear
\left |U\left(\frac{\omega+i V_0}{2}\right)\right|^2  &=&
~\frac{\pi^2}{ \sinh(\pi \omega /\beta) \sinh(\pi \omega \beta)}
\nn \\ &=&
4 \pi^2 ~\sum_{m,n=0}^{\infty}
e^{-2\pi (n+\half) \omega \beta}e^{-2\pi (m+\half) \omega /\beta}
~.
\label{isq}
\eear
Adapting the result in equation (\ref{eqn:review:N/V})
we obtain
\be
\frac {\bar N} {V_p} = \sum_{n,m=0}^{\infty}\int_0^{\infty}
\frac {ds}{s}
\frac{1}{(4\pi s)^{\half{p}}} ~
e^{-\left((n+\half) \beta +(m+\half)/ \beta \right)^2 s}~
\frac{1}{\eta^{D-2} \left ( \frac{is}{2\pi} \right ) }~,
\label{eqn:thermo:N/V}
\ee
where $D-2 = 6(4-V^2)$.

\bibliographystyle{JHEP-like}
\bibliography{dilaton}

\providecommand{\href}[2]{#2}\begingroup\raggedright\begin{thebibliography}{10}

\bibitem{Gutperle:2002ai}
M.~Gutperle and A.~Strominger, {\it Spacelike branes},  {\em JHEP} {\bf 04}
  (2002) 018 [\href{http://arXiv.org/abs/hep-th/0202210}{{\tt
  hep-th/0202210}}].

\bibitem{Sen:2002nu}
A.~Sen, {\it Rolling tachyon},  {\em JHEP} {\bf 04} (2002) 048
  [\href{http://arXiv.org/abs/hep-th/0203211}{{\tt hep-th/0203211}}].

\bibitem{Sen:2002qa}
A.~Sen, {\it Time and tachyon},
  \href{http://arXiv.org/abs/hep-th/0209122}{{\tt hep-th/0209122}}.

\bibitem{Strominger:2002pc}
A.~Strominger, {\it Open string creation by s-branes},
  \href{http://arXiv.org/abs/hep-th/0209090}{{\tt hep-th/0209090}}.

\bibitem{Okuda:2002yd}
T.~Okuda and S.~Sugimoto, {\it Coupling of rolling tachyon to closed strings},
  {\em Nucl. Phys.} {\bf B647} (2002) 101--116
  [\href{http://arXiv.org/abs/hep-th/0208196}{{\tt hep-th/0208196}}].

\bibitem{Chen:2002fp}
B.~Chen, M.~Li and F.-L. Lin, {\it Gravitational radiation of rolling tachyon},
   {\em JHEP} {\bf 11} (2002) 050
  [\href{http://arXiv.org/abs/hep-th/0209222}{{\tt hep-th/0209222}}].

\bibitem{Sen:2002an}
A.~Sen, {\it Field theory of tachyon matter},  {\em Mod. Phys. Lett.} {\bf A17}
  (2002) 1797--1804 [\href{http://arXiv.org/abs/hep-th/0204143}{{\tt
  hep-th/0204143}}].

\bibitem{Sen:2002in}
A.~Sen, {\it Tachyon matter},  {\em JHEP} {\bf 07} (2002) 065
  [\href{http://arXiv.org/abs/hep-th/0203265}{{\tt hep-th/0203265}}].

\bibitem{Constable:2003rc}
N.~R. Constable and F.~Larsen, {\it The rolling tachyon as a matrix model},
  \href{http://arXiv.org/abs/hep-th/0305177}{{\tt hep-th/0305177}}.

\bibitem{Larsen:2002wc}
F.~Larsen, A.~Naqvi and S.~Terashima, {\it Rolling tachyons and decaying
  branes},  {\em JHEP} {\bf 02} (2003) 039
  [\href{http://arXiv.org/abs/hep-th/0212248}{{\tt hep-th/0212248}}].

\bibitem{Mukhopadhyay:2002en}
P.~Mukhopadhyay and A.~Sen, {\it Decay of unstable d-branes with electric
  field},  {\em JHEP} {\bf 11} (2002) 047
  [\href{http://arXiv.org/abs/hep-th/0208142}{{\tt hep-th/0208142}}].

\bibitem{Maloney:2003ck}
A.~Maloney, A.~Strominger and X.~Yin, {\it S-brane thermodynamics},
  \href{http://arXiv.org/abs/hep-th/0302146}{{\tt hep-th/0302146}}.

\bibitem{Lambert:2003zr}
N.~Lambert, H.~Liu and J.~Maldacena, {\it Closed strings from decaying
  d-branes},  \href{http://arXiv.org/abs/hep-th/0303139}{{\tt hep-th/0303139}}.

\bibitem{Gaiotto:2003rm}
D.~Gaiotto, N.~Itzhaki and L.~Rastelli, {\it Closed strings as imaginary
  d-branes},  \href{http://arXiv.org/abs/hep-th/0304192}{{\tt hep-th/0304192}}.

\bibitem{Sen:2003bc}
A.~Sen, {\it Open and closed strings from unstable d-branes},
  \href{http://arXiv.org/abs/hep-th/0305011}{{\tt hep-th/0305011}}.

\bibitem{Shiraz}
S.~Minwalla Private communication.

\bibitem{Schomerus:2003vv}
V.~Schomerus, {\it Rolling tachyons from liouville theory},
  \href{http://arXiv.org/abs/hep-th/0306026}{{\tt hep-th/0306026}}.

\bibitem{Gutperle:2003xf}
M.~Gutperle and A.~Strominger, {\it Timelike boundary liouville theory},
  \href{http://arXiv.org/abs/hep-th/0301038}{{\tt hep-th/0301038}}.

\bibitem{McGreevy:2003kb}
J.~McGreevy and H.~Verlinde, {\it Strings from tachyons: The c = 1 matrix
  reloated},  \href{http://arXiv.org/abs/hep-th/0304224}{{\tt hep-th/0304224}}.

\bibitem{Klebanov:2003km}
I.~R. Klebanov, J.~Maldacena and N.~Seiberg, {\it D-brane decay in
  two-dimensional string theory},
  \href{http://arXiv.org/abs/hep-th/0305159}{{\tt hep-th/0305159}}.

\bibitem{McGreevy:2003ep}
J.~McGreevy, J.~Teschner and H.~Verlinde, {\it Classical and quantum d-branes
  in 2d string theory},  \href{http://arXiv.org/abs/hep-th/0305194}{{\tt
  hep-th/0305194}}.

\bibitem{Myers:1987fv}
R.~C. Myers, {\it New dimensions for old strings},  {\em Phys. Lett.} {\bf
  B199} (1987) 371.

\bibitem{Polchinski:1998rq}
J.~Polchinski, {\it String theory. vol. 1: An introduction to the bosonic
  string}, . Cambridge, UK: Univ. Pr. (1998) 402 p.

\bibitem{Fateev:2000ik}
V.~Fateev, A.~B. Zamolodchikov and A.~B. Zamolodchikov, {\it Boundary liouville
  field theory. i: Boundary state and boundary two-point function},
  \href{http://arXiv.org/abs/hep-th/0001012}{{\tt hep-th/0001012}}.

\bibitem{Teschner:2000md}
J.~Teschner, {\it Remarks on liouville theory with boundary},
  \href{http://arXiv.org/abs/hep-th/0009138}{{\tt hep-th/0009138}}.

\end{thebibliography}\endgroup

\end{document}